\documentclass[reprint,aps,prl,superscriptaddress,
amsmath,amssymb]{revtex4-1}
\usepackage{mathtools}
\usepackage{graphicx}
\usepackage{booktabs,multirow}
\usepackage{braket}
\usepackage{textcomp}
\usepackage{yfonts}
\usepackage[
  bookmarks=false,
  colorlinks,
  linkcolor=blue,
  urlcolor=blue,
  citecolor=blue,
  plainpages=false,
  pdfpagelabels,
  final,
  breaklinks=true
]{hyperref}

\begin{document}
\newcommand{\pd}[2]{\frac{\partial #1}{\partial #2}} 
\newcommand{\td}[2]{\frac{d #1}{d #2}} 

\newcommand{\bs}{\boldsymbol}
\newcommand{\bt}{\textbf}
\newcommand{\sech}{\text{sech}}
\newcommand{\erfc}{\text{erfc}}
\newcommand{\bse}{\begin{subequations}}
\newcommand{\ese}{\end{subequations}}
\newcommand{\im}{\text{i}}
\newcommand{\ud}[0]{\mathrm{d}}
\newcommand{\norm}[1]{\left\lVert#1\right\rVert}

\graphicspath{{figures/},{../figures/}} 
\allowdisplaybreaks

\title{Modal Majorana sphere and hidden symmetries of structured-Gaussian beams}

\author{R. Guti\'{e}rrez-Cuevas}
\email{rodrigo.gutierrez-cuevas@fresnel.fr}
\affiliation{The Institute of Optics, University of Rochester, Rochester, NY 14627, USA}
\affiliation{Center for Coherence and Quantum Optics, University of Rochester, Rochester, NY 14627, USA}
\affiliation{Aix Marseille Univ, CNRS, Centrale Marseille, Institut Fresnel, UMR 7249, 13397 Marseille Cedex 20, France}
\author{S. A. Wadood}
\affiliation{The Institute of Optics, University of Rochester, Rochester, NY 14627, USA}
\affiliation{Center for Coherence and Quantum Optics, University of Rochester, Rochester, NY 14627, USA}
\author{A. N. Vamivakas}
\affiliation{The Institute of Optics, University of Rochester, Rochester, NY 14627, USA}
\affiliation{Center for Coherence and Quantum Optics, University of Rochester, Rochester, NY 14627, USA}
\affiliation{Department of Physics, University of Rochester, Rochester, NY 14627, USA}
\affiliation{Materials Science, University of Rochester, Rochester, NY 14627, USA}
\author{M. A. Alonso}
\email{miguel.alonso@fresnel.fr}
\affiliation{The Institute of Optics, University of Rochester, Rochester, NY 14627, USA}
\affiliation{Center for Coherence and Quantum Optics, University of Rochester, Rochester, NY 14627, USA}
\affiliation{Aix Marseille Univ, CNRS, Centrale Marseille, Institut Fresnel, UMR 7249, 13397 Marseille Cedex 20, France}

\date{\today}

\begin{abstract}
Structured-Gaussian beams are shown to be fully and uniquely represented by a collection of points (or constellation) on the surface of the modal Majorana sphere, providing a complete generalization of the modal Poincar\'e sphere to higher-order modes. The symmetries of this Majorana constellation translate into invariances to astigmatic transformations, giving way to continuous or quantized geometric phases. 
 The experimental amenability of this system is shown by verifying  the existence of both these symmetries and geometric phases.
\end{abstract}


\maketitle


\emph{Introduction.} 
The term ``structured light''  refers to light fields  
with 
nontrivial and interesting amplitude, phase and/or polarization distributions.  A large body of work has been devoted to the production of structured light fields, 
leading to the development of new technologies and the improvement of existing ones 
\cite{andrews2008structured,rubinsztein-dunlop2016roadmap}. Perhaps the best-known example of structured light corresponds to beams carrying orbital angular momentum, used extensively in applications ranging from quantum optics to micromanipulation 
\cite{yao2011orbital,andrews2012angular}. 

\begin{figure}
\centering
\includegraphics[width=.9\linewidth]{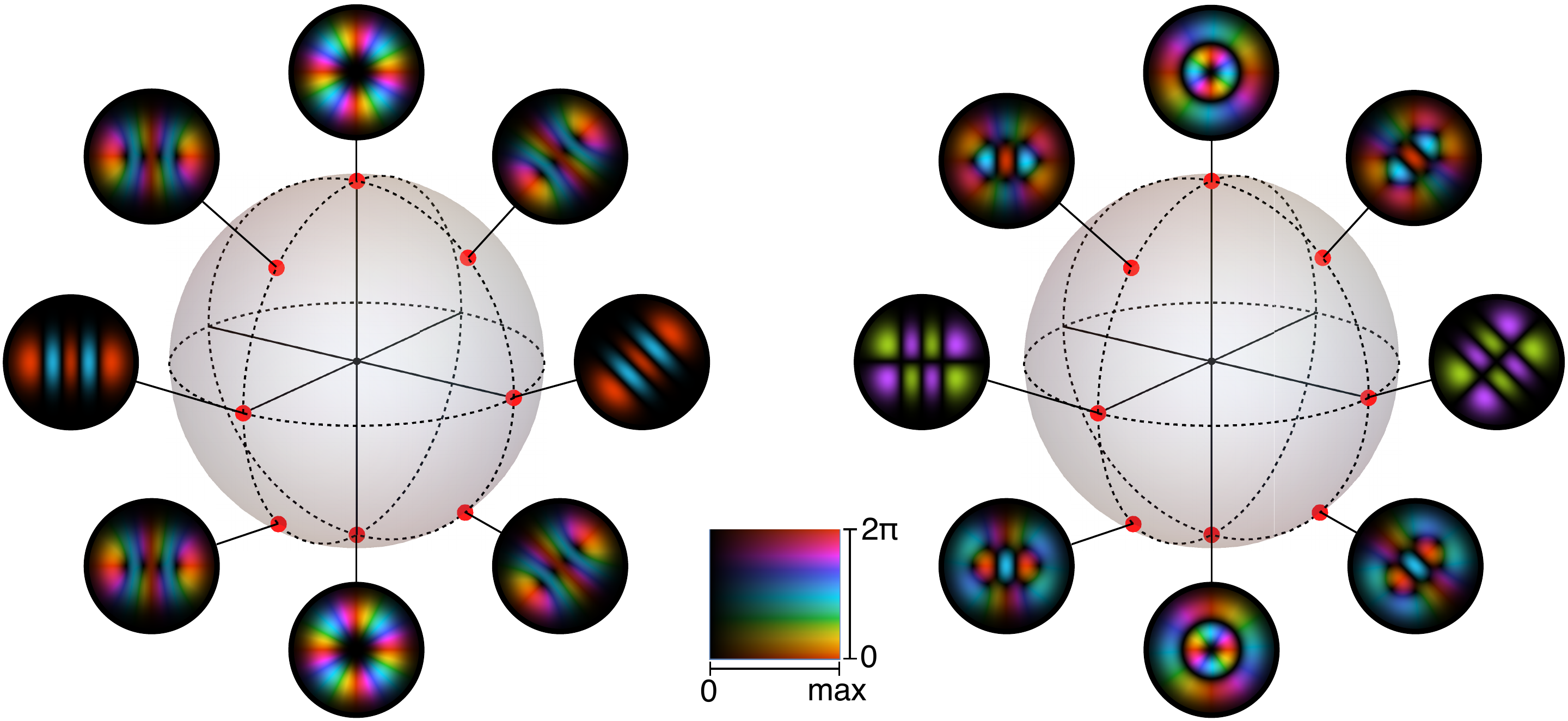}
\caption{\label{fig:MPS4} Modal Poincar\'e sphere for $N=4$ and (left) $\ell=4$,  (right) $\ell=2$. Also shown are the intensity distributions with phase coded in hue, for several HLG beams with their corresponding modal spot. The same color coding is used for all subsequent complex field plots.}
\end{figure}

The current work focuses on the subclass of structured beams referred to as structured-Gaussian (SG) beams 
\cite{dennis2017swings,alonso2017ray,malhotra2018measuring,
dennis2019gaussian}. These solutions to the 
paraxial wave equation have the property of being self-similar, meaning that their intensity profile remains invariant upon propagation up to a scaling factor. SG beams include the well-known Laguerre-Gauss (LG) and Hermite-Gauss (HG) beams 
\cite{siegman1986lasers},
which have been the subject of extensive research and are used for modal decompositions in many applications, such as mode-sorting and subdiffraction localization 
\cite{berkhout2010efficient,zhou2017sorting,gu2018gouy,
tsang2016quantum}. 
LG and HG beams belong themselves to a broader class of SG beams known as the generalized Hermite-Laguerre-Gauss (HLG) modes 
\cite{abramochkin2004generalized,gutierrez-cuevas2019generalized}, which can be obtained from either HG or LG beams by using appropriate pairs of cylindrical lenses (astigmatic transformations) 
\cite{allen1992orbital}. These modes can be represented as points over the surface of a modal Poincar\'e sphere (MPS)
\cite{padgett1999poincare,calvo2005wigner,habraken2010universal} as shown in Fig.~\ref{fig:MPS4}. 
This representation led to the insight that these beams can acquire a geometric phase upon a series of astigmatic transformations~
\cite{pancharatnam1956generalized,berry1984quantal,
enk1993geometric,galvez2003geometric,malhotra2018measuring}. 

The MPS representation of HLG modes
reveals their inherent group structure and transformation properties. There have been generalizations of this construction that mix modal structure and polarization 
\cite{milione2011higher}. However, no generalization 
has been provided for the infinitely more vast family of SG beams of any order (with the exception of a ray-based treatment that is asymptotically valid for fields with well-defined caustics features 
\cite{alonso2017ray,malhotra2018measuring,dennis2019gaussian}).
Here, we show that by introducing the Majorana constellation (MC) 
\cite{majorana1932atomi,bengtsson2017geometry}, originally proposed for spin systems but also used in the context of quantum polarization 
\cite{bjoerk2015extremal}, this longstanding problem can be solved.  
More importantly, the MC provides information about symmetries of SG beams that is not evident from their physical appearance, and the resulting sets of transformations under which they are invariant. These transformations, in turn, lead to geometric phases, which are generally quantized. 
SG beams are more amenable to experimental tests than mathematically-analogous physical systems, making them suitable for testing theoretical predictions, such as the first measurements of discretized geometric phases presented here.

\emph{SU(2) structure.} 
The analogy between SG modes and the quantum eigenstates of a two-dimensional isotropic  harmonic oscillator (2DHO) 
\cite{danakas1992analogies,enk1992eigenfunction,
nienhuis1993paraxial,calvo2005wigner,
habraken2010universal,dennis2017swings,dennis2019gaussian} allows the state and operator formalism to be used for the study of classical optical beams and their propagation; using the Schwinger oscillator model 
\cite{sakurai2010modern,dennis2017swings}, they  can be studied with the help of the operators
\cite{nienhuis1993paraxial,simon2000wigner,calvo2005wigner,
sakurai2010modern,dennis2017swings,dennis2019gaussian}
\begin{align}
 \widehat{T}_0 =& \frac{1}{2w^2} (\widehat x^2+\widehat y^2) +\frac{k^2w^2}{8}(\widehat p_x^2 +\widehat p_y^2), \\
\widehat{T}_1 =&\frac{1}{2w^2} (\widehat x^2-\widehat y^2) +\frac{k^2w^2}{8}(\widehat p_x^2 -\widehat p_y^2), \nonumber \\
\widehat{T}_2 =&\frac{1}{w^2} \widehat x \widehat y +\frac{k^2w^2}{4} \widehat p_x \widehat p_y , \quad
\widehat{T}_3 =\frac{k}{2} (\widehat x \widehat p_y - \widehat y \widehat p_x) , \nonumber
\end{align}
where, in the position representation,
$\widehat x \rightarrow x$ and $ \widehat p_x \rightarrow -\im k^{-1} \partial_{x}$
(and similarly for $y$). These operators satisfy the commutation relations of $\mathfrak{su}(2)$ (spin systems),
$
[\widehat{T}_0,\widehat T_j]=0$, $[\widehat T_i,\widehat T_j]= \im \sum_k\epsilon_{ijk} \widehat T_k$,
with  $i,j,k=1,2$ or $3$ and $ \epsilon_{ijk}$ being the Levi-Civita tensor. 
 $\widehat T_0$ plays the role of the 2DHO Hamiltonian so it describes the propagation of SG beams 
\cite{malhotra2018measuring,ozaktas1994fractional}. Note that $\widehat T_0$ can be used instead of the usual Casimir operator $\lVert \widehat{\bt T} \rVert^2=\lVert(\widehat T_1,\widehat T_2, \widehat T_3)\rVert^2$.

We choose the LG beams as a basis, whose field distribution at the focal plane is given by
\begin{multline}
\label{eq:csLG}
\text{LG}_{N,\ell}(\bt r)=\frac{\im^{|\ell|-N}}{w}\sqrt{\frac{2^{|\ell|+1}\left[(N-|\ell|)/2\right]!}{\pi\left[(N+|\ell|)/2\right]!}} e^{-\frac{r^2}{w^2}} \\ \times
 \left( \frac{r}{w}\right)^{|\ell|} e^{\im \ell \varphi} L_{\frac{N-|\ell|}{2}}^{|\ell|}\left( \frac{2r^2}{w^2} \right),
\end{multline}
where the normalization coefficient includes a phase factor to fulfill the Condon-Shortley condition 
\cite{sakurai2010modern,dennis2017swings,dennis2019gaussian}. LG beams can be denoted by the ket  $\ket{N,\ell}$ indicating the total order $N$  (which determines the Gouy phase) and the azimuthal index $\ell$ of the corresponding LG mode. This notation is motivated by the eigenvalue relations
 $2  \widehat{T}_3 \ket{N,\ell}= \ell  \ket{N,\ell}$ and $ 2 \widehat{T}_0 \ket{N,\ell}=(N+1) \ket{N,\ell} $.
The indices $N$ and $\ell$ play the role of the spin quantum numbers with the minor difference that $\ell$ runs from $-N$ to $N$ in steps of two and $N$ takes only integer values.

\emph{The MPS.} 
The modal transformations generated by the operators $\widehat{T}_i$ can be parametrized in terms of Euler angles \cite{goldstein2001classical,sakurai2010modern},
\begin{align}
\label{eq:dispop}
\widehat{D}(\phi,\theta,\chi)=e^{-\im \widehat T_3 \phi}e^{-\im \widehat T_2 \theta}e^{-\im \widehat T_3 \chi}.
\end{align}
 When $\widehat{D}$ acts on the reference LG beams it transforms them into HLG beams  
 \cite{abramochkin2004generalized} 
$\ket{N,\ell; \bt u} =\widehat{D}(\phi,\theta,0)\ket{N,\ell}$,
 where $\bt u =(u_1,u_2,u_3)= (\cos \phi \sin \theta, \sin \phi \sin \theta, \cos \theta)$ is a unit vector 
\cite{dennis2017swings}. (The parameter $\chi$ was set to zero since it only contributes a global phase.) Two angles $\theta$ and $\phi$ are used to label each HLG mode, which can then be represented by a point (or modal spot) on a sphere  
\cite{padgett1999poincare,calvo2005wigner,habraken2010universal}. This is precisely the standard MPS representation, where the poles represent LG beams with opposite vorticity, the equator represents HG beams with different orientations, and the rest of the sphere represents other HLG beams (see Fig.~\ref{fig:MPS4}). 

\emph{Coherent states and Q function.} 
Coherent states for SG beams can be defined as the extremal HLG states $\ket{N,N}$ satisfying $\widehat{T}_+\ket{N,N}=0$ with $\widehat{T}_\pm=\widehat{T}_1 \pm \im \widehat{T}_2$, in  analogy to spin coherent states  
\cite{radcliffe1971some,arecchi1972atomic,perelomov1972coherent}.
These states, denoted as 
$\ket{N;\bt u} = \ket{N,N;\bt u}$,
have intensity profiles that resemble the elliptic classical orbits of the 2DHO   
\cite{pollett1995elliptic}. In the optical context, coherent states are the beams closest to the elliptical ray families that are the basis of a semiclassical description of SG beams 
\cite{alonso2017ray,malhotra2018measuring,dennis2019gaussian} thus providing a bridge between ray and wave (or  classical and quantum) theories.

\begin{figure}
\centering
\includegraphics[width=.99\linewidth]{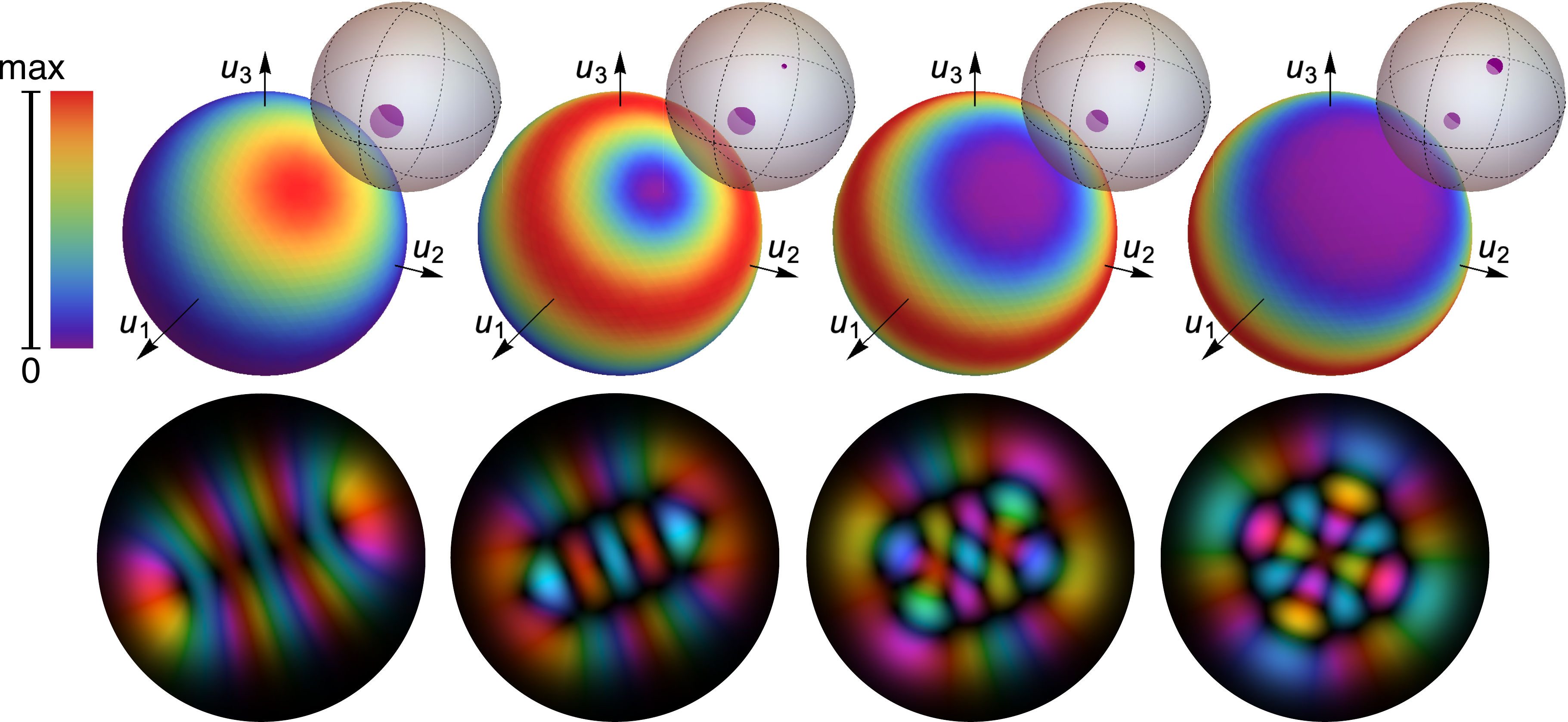}
\caption{\label{fig:hlgMC} 
(First row) Q function and corresponding MC with the size indicating the number of stars. (Second row) Field distribution with hue representing phase for HLG beams of order $N=6$ along $\theta=\pi/4$ and $\phi =\pi/4$ with (from left to right) $l=6,4,2,0$. The same color coding is used for all subsequent plots of Q functions.}
\end{figure}

Coherent states allow defining a Q (or Husimi) function over the reduced phase-space given by the 2-sphere 
\cite{bengtsson2017geometry}. For an arbitrary SG beam  $\ket U$ of total order $N$ this phase space representation is given in terms of its projection onto the coherent states as 
\begin{align}
\text{Q}(\theta, \phi) = \frac{N+1}{4\pi}
\left|\braket{N;\bt u |U}\right|^2.
\end{align}
Figure \ref{fig:hlgMC} shows the Q function for different HLG modes along with their field distribution. The Q function presents a band of high values except for the coherent states for which it is concentrated around a point. 
This ridge outlines a circular path on the sphere which, through the semiclassical description of SG beams, can be used to represent them in terms of rays (with the path's radius encoding the value of $\ell$) \cite{dennis2017swings,alonso2017ray,
malhotra2018measuring,dennis2019gaussian}.

\emph{The modal Majorana sphere (MMS).} Any SG beam with total order $N$ can be decomposed in terms of LG modes as $\ket U= \sum_\ell  c_\ell \ket{N,\ell}$.
The expansion coefficients can then be used to define the $N$th order Majorana polynomial  
\cite{majorana1932atomi,bengtsson2017geometry},
\begin{align}
\label{eq:Mpol}
\psi(\zeta)= \sum_{\ell/2=-N/2}^{N/2}   \sqrt{\binom{N}{\frac{N+\ell}{2}}}  \,c_\ell^*\, \zeta^\frac{N-\ell}{2}.
\end{align}
The zeros of this polynomial map onto those of the Q function through  a stereographic projection $\zeta=\tan ( \theta/2 )\exp (\im \phi )$. Since a polynomial is uniquely determined by its roots, a SG beam is uniquely represented by $N$ points (called stars) on the sphere's surface: the MC
\cite{majorana1932atomi}. (If the number of roots is less than $N$ then  the remaining roots are at infinity, leading to stars at the south pole.) 

\begin{figure}
\centering
\includegraphics[width=.99\linewidth]{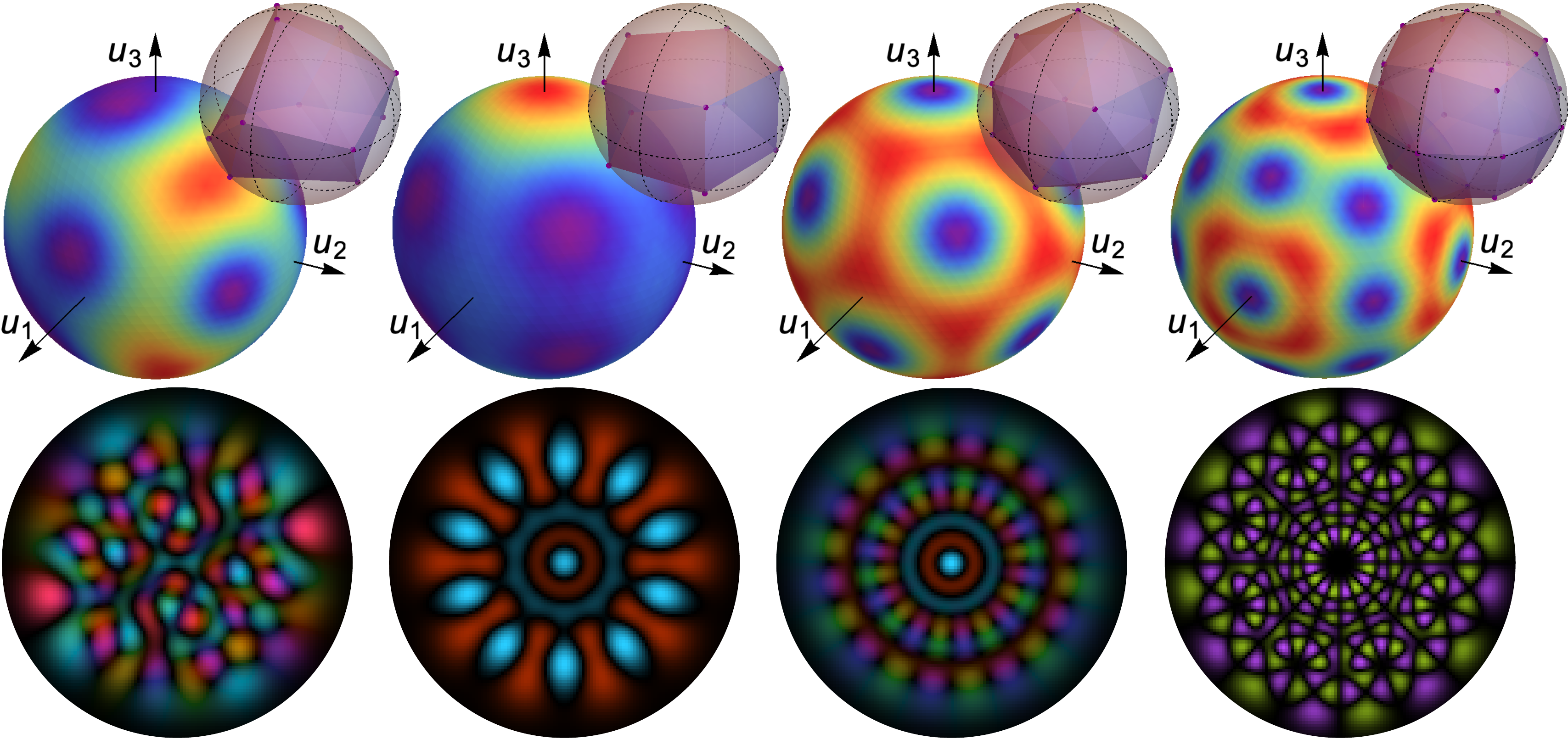}
\caption{\label{fig:MCexamples} (First row) Q function and corresponding MC. (Second row) Field distribution with the phase encoded in hue for four SG beams. The MCs were chosen as (from left to right) arbitrary with no symmetry, a pentagonal prism, an icosahedron, and  a disdyakis dodecahedron. }
\end{figure}

For the HLG mode $\ket{N;\bt u}$  the MC is given by $(N-\ell)/2$ stars at $\bt u$ and $(N+\ell)/2$ stars at $-\bt u$. Figure \ref{fig:hlgMC} shows the MC for different HLG beams. Note that, given our convention, there are more stars at the antipodal point than at the modal spot used in the MPS to represent a HLG beam. Particularly for a coherent state $\ket{N;\bt u}$ all the stars are located at $- \bt u$. 
Other examples of generic SG beams are presented in 
Figure \ref{fig:MCexamples} to show that, unlike the MPS, the MMS represents uniquely any SG beam.

\emph{Hidden symmetries.}  The transformations generated by $\widehat T_j$ act as rotations on the MMS: $\exp (-\im \Omega \bt u \cdot \widehat{\bt T})$ corresponds to a $\Omega$ rotation of the MC around the direction $\bt u$. Therefore, any rotational symmetry of the MC indicates invariance against a modal transformation 
\cite{bacry1974orbits,sakurai2010modern,bengtsson2017geometry}. 
In spin systems, where all axes have similar interpretations, all symmetries have similar physical meaning. 
This is not the case, however, for the optical beams studied here. Rotations generated by $\widehat{T}_3$ correspond to physical rotations of the beam's transverse distribution, and therefore the symmetries of the MC along this axis are evident from the beam's physical appearance (e.g. LG beams are rotationally symmetric since their MC, composed only of stars at the poles, is invariant to rotations along the $u_3$ axis). On the other hand, 
the rotations generated by $\widehat{T}_1$ and $\widehat{T}_2$ correspond to antisymmetric fractional Fourier transformation (fFT), along the Cartesian axes for $\widehat{T}_1$, and along a coordinate system rotated by 45\textdegree{} for $\widehat{T}_2$ (
the latter sometimes referred to as gyration 
\cite{alieva2007orthonormal}).
HG beams are invariant to antisymmetric fFTs because their MC lies on the $u_1$ axis, a symmetry that is not evident from their physical appearance. 
More generally, any symmetry that is not aligned with the $u_3$ axis is not evident from the beam's distribution and thus is ``hidden'' in the abstract representation provided by the MC. 

\begin{figure}
\centering
\includegraphics[width=1.\linewidth]{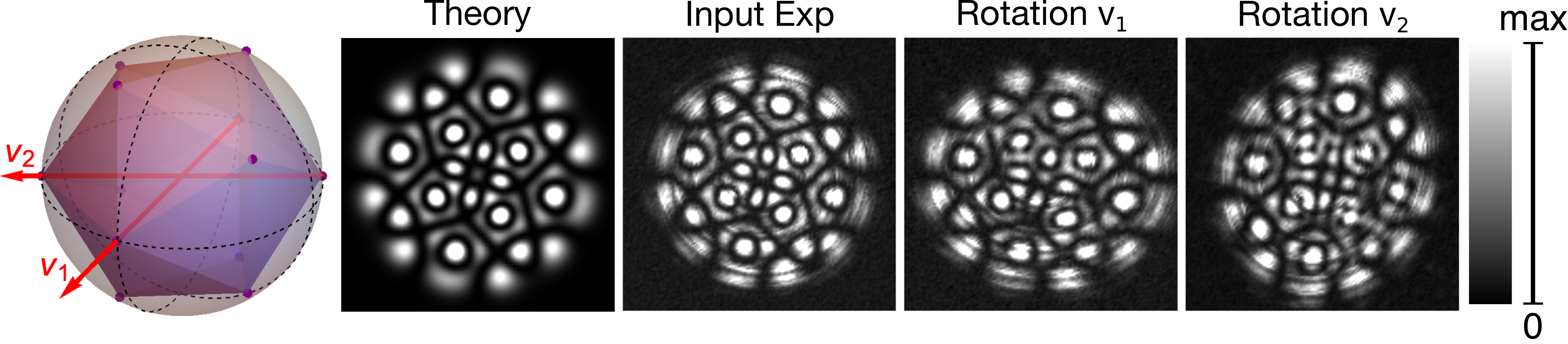}
\caption{\label{fig:ico} (Left to right) Icosahedral MC with four stars laying on the equator, corresponding theoretical and experimental intensity distributions before and after a rotation of the MC along the marked axes.
The theoretical intensity distribution is shown only once since it is the same for all three cases. The same gray scale is used for subsequent intensity plots.}
\end{figure}

The MCs in Fig.~\ref{fig:MCexamples} were chosen to have different symmetries and can be visualized as the vertices of solids.   
As mentioned earlier, only the symmetries of the MC along $u_3$ are evident from the beam's profile. Interestingly, $t$-fold symmetry for the MC corresponds to $2t$-fold symmetry for the beam, 
due to a factor of two connecting rotations in the MMS and in physical space. 
 Different orientations of the MC are obtained by applying the operator in Eq.~(\ref{eq:dispop}) with all three angles being generally nonzero (see Supplemental Material), leading to markedly different beam profiles (e.g. the beams in Figs.~\ref{fig:MCexamples} and \ref{fig:ico} for an icosahedral MC) which nonetheless share the same number and distribution of symmetries. 
The fact that such dissimilar beams with different apparent geometries are connected through simple astigmatic transformations is as surprising as the initial realization that an LG beam can be obtained from an HG beam with just a couple of cylindrical lenses 
\cite{allen1992orbital}.
 
\begin{figure}
\centering
\includegraphics[width=.9\linewidth]{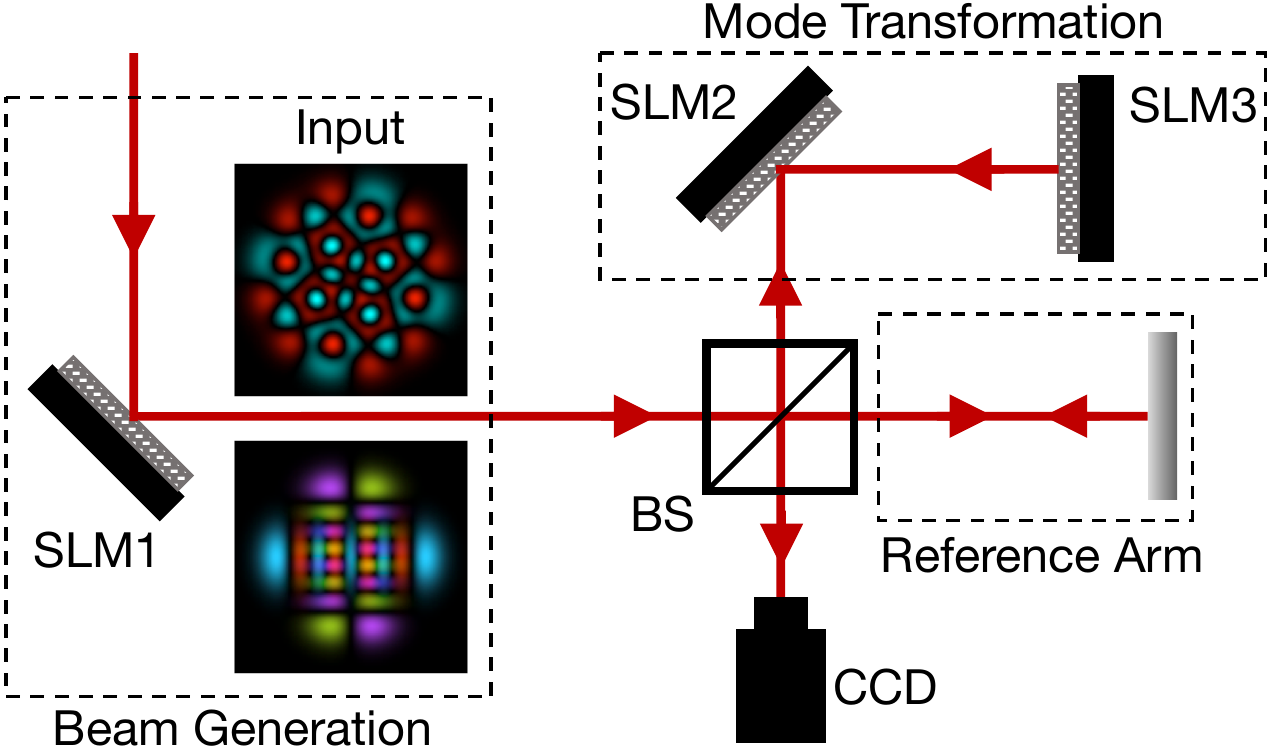}
\caption{\label{fig:exp} Simplified experimental setup. The input SG beams are generated by illuminating SLM1 with a collimated laser beam with $\lambda = 795 \text{nm}$ polarized along the SLM's preferred axis. The beam is then relayed through a 4f system (not shown) to the pair of SLMs that perform the mode transformation. After the mode transformation the beam is relayed to the CCD for detection. When performing interferometric measurements, the input beam is also sent to a reference arm by using a beam splitter and then recombined with the transformed beam. }
\end{figure}

The hidden symmetries can be verified experimentally 
by comparing the beams' profile before and after the corresponding transformation.
The optical setup shown in Fig.~\ref{fig:exp} allows generating with a single spatial light modulator (SLM) the SG beam corresponding to any MC, even for high $N$ \cite{arrizon2007pixelated}. Moreover, any modal transformation represented by a direction $\bt u$ and angle $\Omega$ could be performed with a combination of generalized lenses (implemented with SLMs), and rotators (e.g. dove prisms)
\cite{rodrigo2006optical}. For analogous quantum systems, generating and transforming states corresponding to arbitrary MC is generally not possible.
For example, consider the MC in Fig.~\ref{fig:ico} corresponding to the corners of an icosahedron,  oriented so that the two symmetry axes (red arrows) lie within the equatorial plane. 
The corresponding modal transformations are antisymmetric fFTs along rotated directions, which can be implemented  with two SLMs as shown in Fig.~\ref{fig:exp}
\cite{rodrigo2009programmable,malhotra2018interferometric,malhotra2018measuring}. The powers of the generalized lenses along the rotated directions are set as 
\bse
\begin{align}
p_i^{(\text{SLM2})}=& [1-\cot(\alpha_i/2)/2]/z, \\
p_i^{(\text{SLM3})}=& 2(1-\sin \alpha_i)/z ,
\end{align}
\ese
with $i=x,y$, and where $z$ is the distance between the SLMs, set to $600~\textrm{mm}$ in our setup. By rotating these lenses by an angle $\beta$ any fFT along any axes can be performed. Using SLMs allows the simple tuning of $\alpha_i$ and $\beta$. In particular, when $\alpha_y=-\alpha_x$ the transformation corresponds to an antisymmetric fFT with the axes rotated by $\beta$, which causes a rotation of the MC along axes within the equatorial plane. Figure \ref{fig:ico} shows the intensity distribution of a beam with an icosahedral MC before and after rotations by $2\pi/5$ around the axes $\text{v}_1$ ($\alpha_y=-\alpha_x= \pi/5$ and $\beta =0$)  and $\text{v}_2$ ($\alpha_y=-\alpha_x= \pi/5$ and $\beta =-\text{arctan}\left[(\sqrt{5}-1)/2\right]$); note the difference by a factor of two between the value of the parameters $\alpha_j$ and $\beta$ and the rotation angle and direction.
Except for some slight stretching (caused by a slight mismatch between the beam width and the fFT system's eigenwidth $w=(2\lambda z/\pi)^{(1/2)}$) there is excellent agreement between theory and experiment, showing the beam's invariance to both transformations and its hidden symmetries.

\emph{Discrete geometric phases.} The invariance represented by a rotational symmetry is valid only up to a global phase. For HLG beams this global phase is related to a geometric phase acquired during a cyclic mode transformation 
\cite{pancharatnam1956generalized,berry1984quantal,enk1993geometric,
galvez2003geometric,malhotra2018measuring}, where  the continuous symmetry leads to a phase that varies continuously with the rotation angle. In the general case, however, discrete symmetries lead to discrete (or quantized) values for the geometric phase. 
The MC has been used to study geometric phases given an independent (non-unitary)  evolution of the stars for quantum polarization and spin systems \cite{hannay1998berry,tamate2011bloch,bruno2012quantum,
liu2014representation,ogawa2015observation}. 
We now 
provide a simple formula for the phase acquired by rigid rotations (where the stars do not necessarily trace closed loops) in terms of the symmetry properties of the MC,
and verify it experimentally. 

\begin{figure}
\centering
\includegraphics[width=.9\linewidth]{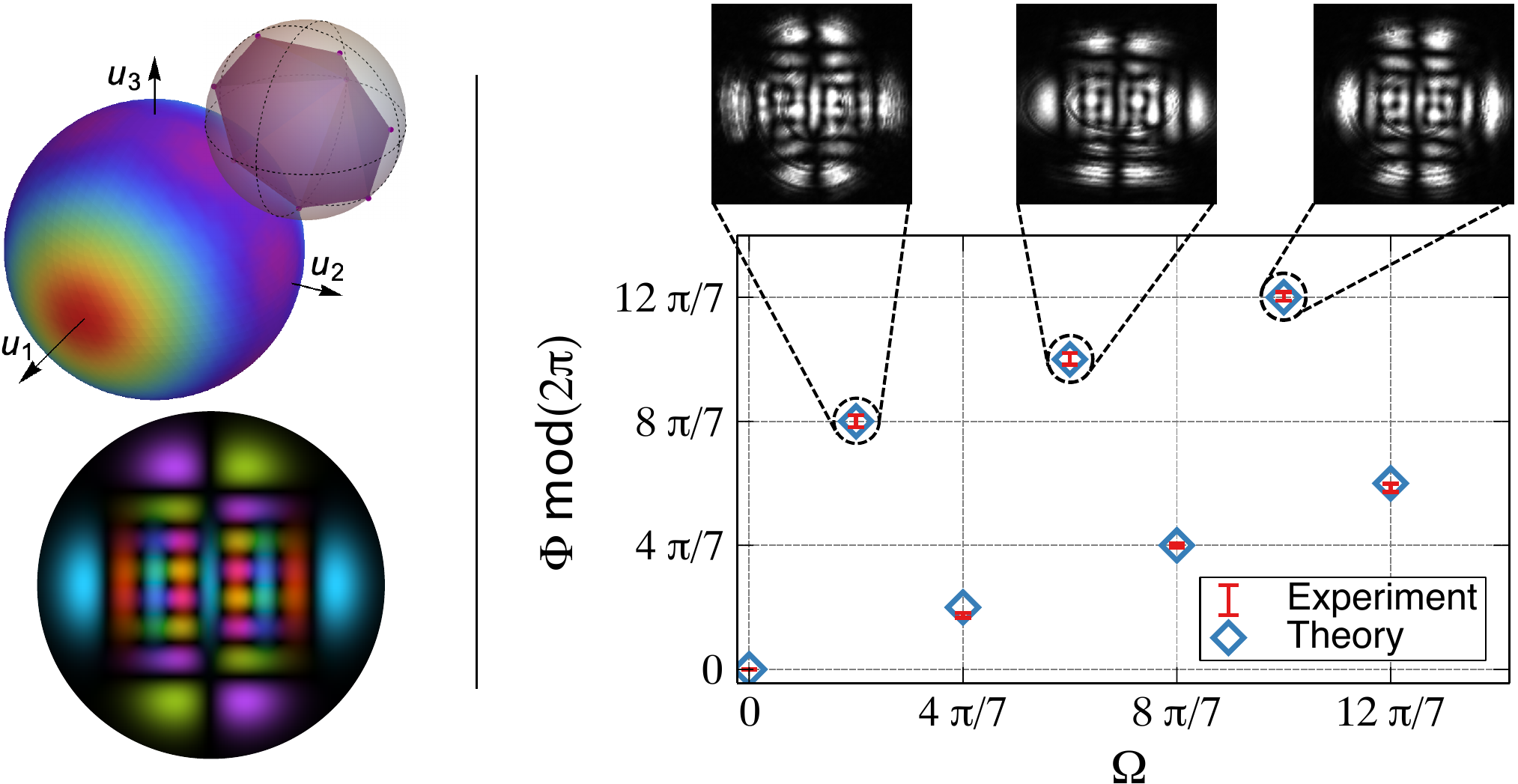}
\caption{\label{fig:disc} (Left) Q function, MC and field distribution for a SG beam with a discrete phase along the $u_1$ axis varying in steps of $8\pi/7$. (Right)  Theoretical and measured geometric phases gained by the beam for different values of the rotation angle along with the intensity distribution of the transformed beam for three rotation angles. }
\end{figure}

Consider a MC with symmetry of order $t$ (i.e.~the smallest rotation leaving the MC invariant is $\Omega = 2\pi/t$) around the direction of $\bt u$. Due to the invariance of the beam, we have $\exp (-\im \Omega \bt u \cdot \widehat{\bt T} )\ket{U}= \exp (-\im \Phi ) \ket{U}$
where 
\begin{align}
\label{eq:discphi}
\Phi =& \pm 2\pi \left( \frac{ N - 2s_\pm}{2t}\right) \; \text{mod}(2\pi), 
\end{align}
with $s_\pm$ being the number of stars at $\pm \bt u$ (see Supplemental Material for the proof). This formula involves only information in the MC. Since we considered the smallest possible rotation, the geometric phase increases in steps of $\Phi$ as the angle of rotation increases in steps of $2\pi/t$. 
The optical setup allows verifying this geometric phase by adding a reference arm (see Fig.~\ref{fig:exp}) in a coupled Michelson-interferometer configuration (see SM for specific details).
Equation (\ref{eq:discphi}) not only permits the simple computation of the geometric phase arising from cyclic transformations, but it also provides a way to design SG beams that gain specific amounts of geometric phase after a given transformation. For example,
 Fig.~\ref{fig:disc} shows a SG beam designed to gain a geometrical phase of $8\pi/7$ after a rotation of $2\pi/7$ of the corresponding  MC around the $u_1$ axis.
This figure also shows the measured geometric phase gained after each rotation of $2\pi/7$, showing good agreement with theory. The intensity distributions shown in Fig.~\ref{fig:disc} demonstrate again that rotational symmetries of the MC translate into invariance under specific astigmatic transformations. We believe this is the first experimental verification of the quantized nature of the geometric phase when considering the cyclic evolution of a complex~MC.

\emph{Conclusions.} The MMS provides a simple geometric construction to represent any SG beam, thus 
fully generalizing the MPS for higher-order modes. 
It can be used as an intuitive beam design tool that reveals the beam's hidden symmetries. This construction also shows that the geometric phases arising from cyclic modal transformations of general SG beams can be discrete.  
Note that the geometric phase can be defined in a consistent way for noncyclic evolutions as the argument of $\bra{U} \exp (-\im \Omega \bt u \cdot \widehat{\bt T} )\ket{U}$. This phase presents a highly nonlinear behavior in $\Omega$ (similar to the case treated in 
\cite{tamate2011bloch}) that interpolates between the discrete values defined by cyclic transformations. 

SG beams have applications in micromanipulation, information transfer, imaging and metrology \cite{yao2011orbital,andrews2012angular,siegman1986lasers,
berkhout2010efficient,zhou2017sorting,gu2018gouy,
tsang2016quantum,bouchard2017quantum}, so results that elucidate their underlying structure and properties can have an important impact. 
Particularly, since for SG beams the axes in the Majorana representation correspond to physical rotations and astigmatic transformations, NOON-like states can be used to measure misalignments and asymmetric deviations in the curvature of optical elements.
SG beams also provide a platform for testing theoretical predictions relating to analogous quantum mechanical systems, given the high level of control over their generation and transformation. Moreover, the results presented here can be translated into the quantum regime given that single photons possess a modal structure that can be shaped with SLMs.
Another interesting avenue to be pursued separately is the study of the most symmetric MC 
\cite{conway1996packing,saff1997distributing,
aulbach2010maximally,giraud2010quantifying,bjoerk2015extremal} and the resulting beams. These MCs are often related to the ``most quantum'' states. In the optical context these are the ``least ray-like'' beams thus it will be interesting to test whether a ray-based description is still valid.  

The authors acknowledge M.~R.~Dennis, L.~L.~S\'anchez-Soto and E.~Pisanty for useful discussions. The work by R.G.C. and M.A.A. was supported by the National Science Foundation (PHY-1507278) and the Excellence Initiative of Aix-Marseille University - A*MIDEX, a French ``Investissements d'Avenir'' programme. A.N.V and S.A.W acknowledge support from the DARPA YFA \# D19AP00042.

R.~G.-C.~and S.~A.~W.~contributed equally to this work.


%

\end{document}